\documentclass[pre,aps,eqsecnum,showpacs,twocolumn]{revtex4}
\newcommand{\nothesis}[1] {#1}
\newcommand{\nopaper}[1] {}

\newcommand{\Poincare}{Poincar\'{e}}
\newcommand{\Henon}{H\'{e}non}
\newcommand{\eqn}[1] {Eq. (#1)}

\newcommand{\avgdotphi}{<\!\!\dot{\Phi}_0\!\!>}

\newcommand{\transop}{\mathbf{T}}
\newcommand{\swapop}[1]{\mathbf{E}_{#1}}

\newcommand{\Mock}{A}

\newcommand{\amatrix}[1] {{\mathbf #1}}

\newcommand{\SPIN}{{\bf S}}
\newcommand{\spin}{S}
\newcommand{\monod}{\amatrix{M}}
\newcommand{\trace} {{\mathrm{Tr} \,}}
\newcommand{\norm}[1]{\mid#1\mid}

\nothesis{
  
}

\input epsf
\newcommand{\Section}[1]{\section{#1}}
\begin{document}

\title{Classical Mechanics of a Three Spin Cluster}
\author{P. A. Houle and C. L. Henley}
\affiliation{Dept. of Physics, Clark Hall, Cornell University, Ithaca NY 14853}

\begin{abstract}
A cluster of three spins with single-axis anisotropic exchange
coupling exhibits a range of classical behaviors,  ranging from
regular motion at low and high energies to chaotic motion at
intermediate energies.  A change of variable makes it possible to
isolate total angular momentum around the z-axis (a conserved quantity)
and it's associated cyclic variable from two non-trivial degrees
of freedom.  This clarifies the interpretation of \Poincare\ sections,
and causes permutation symmetries of the system to manifest as
rotation symmetries in the new coordinates.

The three-spin system has four families of periodic orbits (multiple
instances of each orbit exist because of permutation symmetry.)
Analysis of spin waves predicts their periods in the low energy
(antiferromagnetic) and high energy (ferromagnetic) limits and also
can be used to determine the stability properties of certain
orbits at intermediate $E=-1.$  The system also undergoes interesting
changes in the topology of the energy surface along particular curves
in energy-parameter space,  for instance,  when two pieces of the
energy surface surrounding the two antiferromagnetic fixed points
coalesce.  \Poincare sections produced with a 3-d graphic technique
(described in the Appendix) illustrate the symmetries of the system
and illustrate the transition to chaos at low energies.

The three spin system turns out to have similarities with the
Anisotropic Kepler Problem (AKP) and the \Henon-Heiles Hamiltonian.
An appendix discusses numerical integration techniques for spin
systems.  The quantum
manifestations of the structures found in this paper are discussed
in  [P. A. Houle, N. G. Zhang, C. L. Henley, Phys.\ Rev.\ B \textbf{60}
15179 (1999)].


\end{abstract}
\pacs{03.65.Sq, 05.45.+b}

\maketitle


\Section{Introduction}
        This paper is a study of the classical mechanics of a cluster
of three spins with single-axis anisotropic exchange coupling;  the
Hamiltonian of the system is

\begin{equation}
H = J \left[ \sum_{1\le i < j \le 3}\; {\SPIN}_i \cdot {\SPIN}_{i+1} - \sigma \spin_{iz} \spin_{(i+1)z} \right],
\label{eq:hamiltonian}
\end{equation}

We can set $J=1$ without losing 
generality (even the sign is arbitrary for our study of the dynamics). 
@@
\eqn{\ref{eq:hamiltonian}} is a model for spin clusters in triangular
antiferromagnets such as $\mathrm{NaTiO_2}$ and 
$\mathrm{RbFeCl_3}.$  It was introduced in \cite{Nakamura85}
and is further reported on in 
\cite{Nakamura86,Nakamura86a,Nakamura93}.
In these works,  Nakamura
studied the level statistics of the system,
and compared the behavior of energy level fluctuations as
a function of $\sigma$ in regions where the classical dynamics was
predominantly regular and chaotic.  In a previous work,  we've discussed
how the classical structures of the system are manifested in its quantum
spectrum \cite{Houle98b}.

This system is of intrinsic interest as a comparatively simple
dynamical system possessing high symmetry. We were specifically
motivated by semiclassics, i.e. the relation between classical
dynamics and the eigenstates of a quantum system.  

The quantum manifestations of classical chaos were investigated
not only on the model of \eqn{\ref{eq:hamiltonian}},  
but also in a cluster of two interacting spins 
\cite{Magyari87,Srivastava88,Srivastava90a,Srivastava90b,Srivastava90c,Srivastava91}.  
The method of Bohr-Sommerfeld quantization
has also been applied, but  to single spins \cite{Shankar80,Klein81}.

In recent years, a variety of magnetic molecules have been synthesized
containing clusters of interacting spins~\cite{Sessoli93}. 
Although commonly approximated as a single moment, they do have
internal excitations which cannot be found exactly by exact diagonalization 
(since the Hilbert space is too large).  If the spin cluster
divides into subclusters,  each of which has a moderately long moment, 
these excitations may best be grasped semiclassically. 
The methods of the present paper would be a natural starting point
for such studies, at least for systems with high symmetry.

\eqn{\ref{eq:hamiltonian}} 
conserves total $\spin_z=\sum_{i=1}^3 \spin_{iz}.$  A key aspect of
our approach is a change of variable which separates the system into
two subsystems:  (i) A nonintegrable two-degree of freedom subsystem,  and
(ii) a single degree of freedom subsystem for which total $\spin_z$ is
a conserved momentum variable.  
Our change of variable makes the  \Poincare\ sections comprehensible
[in contrast to prior work~\cite{Nakamura93} that
did not separate subsystems (i) and (ii)], 
and makes it easy to find the  fundamental
periodic orbits of the system.
As in previous work on
\eqn{\ref{eq:hamiltonian}},  we restrict our consideration to the
special case that $\spin_z=0;$  as this paper considers only the classical
mechanics of the system,  we set $|\SPIN|=1$ without losing generality.

An outline of this paper follows: Section \ref{sec:newtile} introduces our change
of variable and shows how the topology of the phase space
and the symmetries of the system appear in the new coordinates.  Section \ref{sec:c3fp}
is a discussion of the fixed points,  invariant manifolds,  and spin waves of the three-spin
system.  Section \ref{sec:c3topology} is about changes in the topology of the energy surface that occur
along certain curves in the $E-\sigma$ parameter space.  Section \ref{sec:c3pos} enumerates the fundamental
periodic orbits of the system and maps the global dynamics of the system with
\Poincare\ sections.  Section \ref{sec:c3conclusion} concludes by
discussing similarities between the the three-spin cluster and the well-studied
Anisotropic Kepler Problem and the \Henon-Heiles Hamiltonian.

\Section{Hexagonal phase space}
\label{sec:newtile}
With three degrees of freedom,  the dynamics of an arbitrary dynamical
system is difficult to visualize and understand.  Fortunately,  the
three-spin cluster conserves
total angular momentum around the z-axis,  so we can separate the problem into
two halves:
(i) an autonomous nonintegrable two degree of freedom Hamiltonian,
and (ii) an trivial single degree of freedom system which is driven by
system (i). Part (i) can  be studied in isolation from part (ii),  however,
since (ii) is driven by (i) we must solve the motion of (i) before we
can solve the motion of (ii).

Section \ref{sec:change} presents a change of variable that separates part (i) from 
part (ii) -- which is necessary to draw useful \Poincare\ sections.  Next, Section \ref{sec:pzsym}
shows how the discrete symmetries of the system appear in the new coordinates.  Then,  
in Section \ref{sec:howtile} we'll show that the $2 \pi$ periodicity
of the spin longitudes $(\phi_i)$ manifests as hexagonal tiling in the
$(\Phi_A, \Phi_B)$ coordinates.

\subsection{A change of coordinates} 
\label{sec:change}

\par

The conventional canonical coordinates for spin are
$(\phi_i,z_i)$ where $\phi_i$ is longitude and $z_i=\spin_{iz}.$
\cite{vanHemmen86}  Spin space includes both position and momentum and
is the complete phase space of a spin system -- the position of the spin
vectors at a moment in time completely describes the system.  The
mapping between $(\phi_i,z_i)$ and spin vector components is 
${\SPIN_i} = \{\spin_{ix},\spin_{iy},\spin_{iz} \}=\{ (1-z_i^2)^{1\over 2} 
\cos \phi_i,(1-z_i^2)^{1 \over 2} \sin \phi_i, z_i\}$.

To isolate total angular momentum around the $z$ axis,  we make
the orthogonal linear transformation
\begin{equation}
\left(
  \begin{array}{c}
    \Phi_0 \\
    \Phi_A \\
    \Phi_B
  \end{array}
\right)=
\mathbf{M}
\left(
  \begin{array}{c}
    \phi_1 \\
    \phi_2 \\
    \phi_3 
  \end{array}
\right), 
\left(
  \begin{array}{c}
    Z_0 \\
    Z_A \\
    Z_B
  \end{array}
\right)=
\mathbf{M}
\left(
  \begin{array}{c}
    z_1 \\
    z_2 \\
    z_3 
  \end{array}
\right), \label{eq:newcoords}
\end{equation}

where

\begin{equation}
\mathbf{M}=
{1 \over \sqrt{6}}
\left(
  \begin{array}{ccc}
    \sqrt{2} & \sqrt{2}  &  \sqrt{2}   \\
    2 & -1 & -1   \\
           0           & \sqrt{3}  & -\sqrt{3}   \\
  \end{array}
\right).
\end{equation}

Because $\Phi_0$ is a cyclic coordinate (does not appear in \eqn{\ref{eq:newHamiltonian}}),
$Z_0$ is conserved: $Z_0$ is proportional to total spin around the $z$-axis and $\Phi_0$ measures the collective
precession of the three spins around the z-axis.  We call the
two degree-of-freedom system consisting of ($\Phi_A$,$\Phi_B$,$Z_A$,$Z_B)$ 
the \emph{reduced system} and the complete system with three degrees
of freedom the \emph{full system.}  As the equations of motion for the
reduced system do not depend on $\Phi_0,$  the evolution of $\Phi_0$
can be ignored when we study the reduced system; $Z_0$ enters only as
a constant parameter of the reduced system.

$Z_0=0$ when total $\spin_z=0$ is zero (as is always the case in this paper), and the Hamiltonian becomes
\begin{eqnarray}
H&=&-{ (1-\sigma) \over 2}
\left(Z_A^2+Z_B^2\right)
+ f_+ f_- \cos
\left(
\sqrt{2} \Phi_B
\right)
\nonumber \\
&&
+ f_0 f_+ \cos
\left( 
 {
 \Phi_B+ \sqrt{3} \Phi_A
 \over \sqrt{2}
 }
\right) 
\nonumber \\
&&
+f_0 f_- \cos
\left( 
 {
 \Phi_B- \sqrt{3} \Phi_A
 \over \sqrt{2}
 }
\right),
\label{eq:newHamiltonian}
\end{eqnarray}
in the new coordinates with
\begin{eqnarray}
f_0&=&\sqrt{1-{2 \over 3} Z_A^2}, \\
f_\pm&=&\sqrt{1-{1 \over 6} \left(Z_A \pm \sqrt{3}Z_B\right)^2}. \label{eq:endNewH}
\end{eqnarray}
%

If we're interested in the evolution 
of $\Phi_0$,  we can study the full system by first finding
a trajectory of the reduced system and then solving the remaining time-dependent
equation of motion for $\Phi_0.$  A periodic orbit of the reduced system
may or may not be a periodic orbit of the full system;  a periodic
orbit of the reduced system is a periodic orbit of the full system
only if $\Phi_0$ changes by an integer multiple of $2 \pi/\sqrt{3}$ per orbit of the reduced
system.

\subsection{Symmetries in $\Phi,Z$ coordinates}

\label{sec:pzsym}

Structures in phase space,  such as invariant manifolds and fundamental periodic orbits,
reflect symmetries of a dynamical system.  Therefore,  it's essential to understand the discrete symmetries of a 
system in order to characterise structures in phase space.

Permutations of the identities of the spins are an important set of
discrete symmetry operations for the three spin cluster.
For instance, if we swap $\SPIN_2$ and $\SPIN_3$ the Hamiltonian 
\eqn{\ref{eq:hamiltonian}} is unchanged.
The three spin cluster is the $N=3$ case of a simplex, in which all of
the spins and the bonds between the spins are interchangeable.  The
associated symmetry group is the permutation group $P_3$ which
requires two generators. The first generator is $\transop$, the
\emph{translation} operator.  $\transop$ maps $\SPIN_1\rightarrow\SPIN_2$, 
$\SPIN_2\rightarrow\SPIN_3$, and $\SPIN_3\rightarrow\SPIN_1.$  Viewed in $(\Phi,Z)$ coordinates,
$\transop$ is a rotation of both the $(\Phi_A,\Phi_B)$ and $(Z_A,Z_B)$
planes by $2 \pi/3$ or
\begin{equation}
\left(
  \begin{array}{c}
    \Phi_0 \\
    \Phi_A \\
    \Phi_B
  \end{array}
\right)=
\mathbf{T}
\left(
  \begin{array}{c}
    \phi_1 \\
    \phi_2 \\
    \phi_3 
  \end{array}
\right), 
\left(
  \begin{array}{c}
    Z_0 \\
    Z_A \\
    Z_B
  \end{array}
\right)=
\mathbf{T}
\left(
  \begin{array}{c}
    z_1 \\
    z_2 \\
    z_3 
  \end{array}
\right), 
\end{equation}
with 
\begin{equation}
\transop=
{1 \over 2}
 \left(
  \begin{array}{ccc}
   2 & 0 & 0 \\
   0 & -1 & - \sqrt{3} \\
   0  & \sqrt{3} & - 1 \\
  \end{array}
 \right).
\end{equation}
the same transformation over $Z.$  The {\it exchange} operators
are a subset of $P_3$:  $\swapop{i}$,  where $i$ is a spin index,
exchanges the other two spins.  $\swapop{1}$ is
\begin{equation}
\left(
  \begin{array}{c}
    \Phi_0 \\
    \Phi_A \\
    \Phi_B
  \end{array}
\right)=
\swapop{1}
\left(
  \begin{array}{c}
    \phi_1 \\
    \phi_2 \\
    \phi_3 
  \end{array}
\right), 
\left(
  \begin{array}{c}
    Z_0 \\
    Z_A \\
    Z_B
  \end{array}
\right)=
\swapop{1}
\left(
  \begin{array}{c}
    z_1 \\
    z_2 \\
    z_3 
  \end{array}
\right),
\end{equation}
with
\begin{equation}
\swapop{1}=
 \left(
  \begin{array}{ccc}
   1 & 0 & 0 \\
   0 & 1 & 0 \\
   0  & 0 & - 1 \\
  \end{array}
 \right).
\end{equation}
in $(\Phi,Z)$ coordinates.  (The same transformation matrix also acts on
the $Z$ coordinates).  Any $E_i$ can be chosen for a second generator -- 
$\transop$ and $\swapop{1}$ are a complete set of generators for $P_3$.

\par

\subsection{The tiling of $\Phi,Z$ coordinates}
\label{sec:howtile}

Unlike the phase space of
particle systems,  which is infinite in extent,  spin space is
compact and periodic in $\phi_i.$  Therefore,  spin trajectories exist which
have no analog
in a particle system.  For instance,  a spin can precess around
the z-axis and come back to its initial position without the
sign of $\dot{\phi}$ ever changing.  Also,  a spin can pass directly
over  the north pole $(z_i \rightarrow 1)$ at which point $\phi_i$ jumps 
discontinuously by $\pi.$  

The transformation \eqn{\ref{eq:newcoords}} also changes the
appearance of the boundaries and connectivity of the phase space;  the periodicity of 
the $\phi$ coordinates causes $(\Phi_A,\Phi_B)$ to be periodic on a hexagonal lattice 
while the condition
$|z_i|<1$ restricts $(Z_A,Z_B)$ to the interior of a hexagon.

\begin{figure}
\centerline{\epsfxsize=3.2in\epsfbox{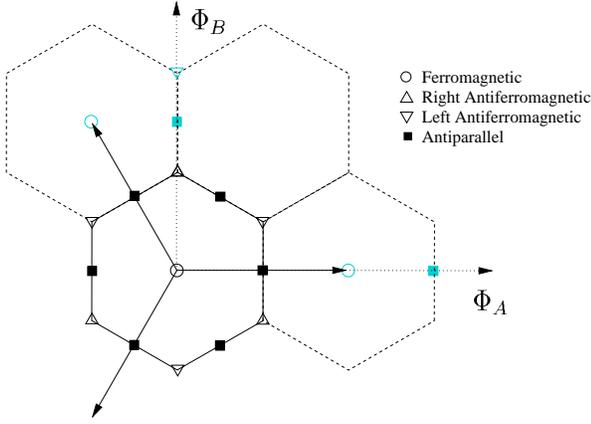}}
\caption{The hexagonal tiling of the $\Phi_A,\Phi_B$ plane.
The arrows are the three lattice vectors.  The primitive unit cell
is solid and copies are dashed.  The following fixed points are
depicted:  (i) the one ferromagnetic (FM) fixed point,  (ii)
the two antiferromagnetic (AFM) fixed points,  and (iii) the
three antiparallel (A) fixed points.}
\label{fig:lattice}
\end{figure}

Because the the $\phi$ coordinates are $2 \pi$ periodic
the transformation $\phi_i \rightarrow \phi_i + 2\pi$ does
not change the state of the system.  This periodicity looks different
in the $(\Phi_A,\Phi_B)$ plane:  the transformation $\phi_1 \rightarrow
\phi_1 + 2 \pi$ maps to to a translation in $(\Phi_A,\Phi_B)$ space of
$({4 \pi \over \sqrt{6}},0).$  Adding $2 \pi$ to $\phi_2$ maps to a
translation of ${2 \pi \over \sqrt{6}}(-1,\sqrt{3})$ 
and adding $2\pi$ to $\phi_3$ maps to a translation of ${2 \pi \over
\sqrt{6}} (-1,-\sqrt{3})$ -- related by the operator $\transop,$ these three vectors form the lattice vectors of a hexagonal lattice,
see Fig. \ref{fig:lattice}.

\par

Since the $z$'s and $Z$'s are related by the same linear transformation
that relates the $\phi$'s and $\Phi$'s,  the domain of valid $(Z_A,Z_B)$
is hexagonal.  
One boundary of the hexagon is where spin $1$ is at the
north pole,  $z_1=1,$ since $z_i \in [-1,1]$.  At that point, $Z_A={\sqrt{6} \over 2}.$  When spin $1$
is at the south pole,  $z_1=-1$ and $Z_A=-{\sqrt{6} \over 2}.$  The
rest of the boundaries can be found by rotating the $z_1=\pm 1$ boundaries
by $\pm 120^{\circ}$ in the $(Z_A,Z_B)$ plane.

A remaining detail is how the $\Phi,Z$ trajectory appears 
as a spin passes through a pole.  For example,  when spin 1 hits the north pole,
the the trajectory strikes the boundary of the 
$(Z_A,Z_B)$ hexagon at $Z_A={\sqrt{6} \over 2}.$  Although there is no
discontinuity in the $z$ or $Z$  coordinates,  the
$\phi_1$ coordinate jumps discontinuously by $\pi$.  
As seen in $(\Phi_A,\Phi_B)$ coordinates,  the coordinate $\Phi_A$
jumps by $2 \pi \over \sqrt{6}.$  As $\dot{z_1}$ changes
discontinuously at this point,  the projection of the trajectory seems
to ``bounce'' off the boundary in $(Z_A,Z_B.)$


\Section{Fixed Points,  invariant manifolds and spin waves}
\label{sec:c3fp}

\subsection{Fixed Points}
\label{sec:c3fpa}
        Certain phenomena of the three spin system,  such as
fixed points,  invariant manifolds and spin waves,  can be
studied without numerical integration. 
Section \ref{sec:c3fpa} concerns the fixed points of the three-spin system in which all
spins lie in the equatorial plane.  Next, Section
\ref{sec:manifolds} describes invariant manifolds of the
system -- two-dimensional subspaces on which the dynamics are reduced
to a single degree of freedom.  Finally, Section \ref{sec:spinWaves}
develops a linear expansion around the fixed points found in Section
\ref{sec:c3fpa} to derive the frequency of spin wave excitations in
their phase-space vicinity.

Fixed points are points in phase space where 
$\dot{\Phi}_j=\dot{Z}_j=0$ for $j \in \{0,A,B\}$  There are three families of
fixed points of the system in which the
spins lie in the equatorial plane ($Z_A=Z_B=0.$)  The locations of these fixed
points in $(\Phi_A,\Phi_B)$ space are plotted on Fig. \ref{fig:lattice}.
The ferromagnetic state ($FM$)  is the global energy maximum ($E=3$) with all three spins pointing together.
 The $FM$ state lies at $(0,0)$ in the $(\Phi_A,\Phi_B)$ plane.  
The two antiferromagnetic states
($AFM_{R}$ and $AFM_{L}$) 
are located at $(0, \pm {2 \sqrt{2} \pi \over 3})$ in $(\Phi_A,\Phi_B)$ plane
and are mirror images of each other.  The antiferromagnetic states are global energy minima ($E=-1.5$) with 
the spins splayed $120^\circ$ apart.  There are also three
\textit{antiparallel} configurations
($A_i$ where $i$ is a spin index $i \in \{1,2,3\}$) where two spins are coaligned while the other
spin (spin $i$) points in the opposite direction;  here $E=-1.$  $A_1$
lies at $(0, \sqrt{2} \pi)$ while $A_2$ and $A_3$ lie at
$(\pm \sqrt{3/2}\pi,\sqrt{1/2}\pi)$ 
in the $(\Phi_A,\Phi_B)$ plane -- the operator $\transop$ transforms
into another by the operator $\transop,$  a $120^\circ$ rotation in the
$(\Phi_A,\Phi_B)$ plane.   The $F$,  $AFM$ and $A$ fixed points are fixed points of both the reduced and full systems.

\par

\subsection{Invariant Manifolds}
\label{sec:manifolds}

If a subspace of the phase space is an invariant manifold,  the time evolution
of the system will remain in that subspace if its initial state lies in that
subspace.  As the invariant manifolds of the reduced
system are two dimensional,  dynamics on the invariant manifolds possess only
a single degree of freedom.  Therefore,  a family of periodic orbits lives on
each invariant manifold,  which are discussed further in 
Section \ref{sec:c3pos}.  The three spin system has two kinds of invariant manifold:  the {\it stationary spin} manifolds and the
{\it counterbalanced} manifolds.  For either kind of manifold,  the motion
of one spin is different from the other two;  with the operator $\transop$ 
one can find three manifolds of each type,  related to one another by a
$120^\circ$ rotation in the $(\Phi_A,\Phi_B)$ or $(Z_A,Z_B)$ planes.

The $\Phi_A=Z_A=0$ subspace is one stationary spin manifold.  On this manifold,  spin 1 lies in the
equatorial plane and remains stationary while the other two spins execute
roughly circular motions in opposite directions.  Motion on the
stationary spin manifolds can be modelled with a
one-spin system:  pointing spin 1 along the $x$-axis,  the constraint
$\Phi_A=Z_A=0$ combined with $\sum S_z=0$ implies that
$\spin_{3x}=\spin_{2x},$ $\spin_{3y}=-\spin_{2y}$
and ${\spin}_{3z}=-{\spin}_{3z}.$  In spin vector form,  the
reduced Hamiltonian is
\begin{equation}
H_{ss}=2 \spin_x+\spin^2_x-\spin^2_y-(1-\sigma) \spin^2_z, \label{eq:ssSingle}
\end{equation}
where $\SPIN=\SPIN_2.$

\par

The other family of invariant manifolds are the counterbalanced manifolds in
which two spins move together in a direction opposite to the other spin;
the counterbalanced manifold with spin one the special spin is the
subspace $\Phi_B=0, Z_B=0.$  With the arbitrary choice of $\Phi_0=0,$ the
following constraints
apply: $\SPIN_2=\SPIN_3$,  $\spin_{1y}=-2 \spin_{2y}$ and $\spin_{1z}=-2\spin_{2z}.$
The remaining constraint, on $\spin_{2x}$,  is determined by the total length constraint $\norm{\SPIN_1}=1,$  which implies
$\spin_{2x}=\left(+\sqrt{3 +\spin_{1x}^2}\right)/2.$  
With $\SPIN=\SPIN_1$ the Hamiltonian reduces to

\begin{equation}
H= J \left[ 1 + \spin_x \sqrt{3+\spin_x^2} - \spin_y^2 - \spin_z^2
+ \sigma {3 \over 4} \spin_z^2\right].
\end{equation}

\subsection{Spin Waves}
\label{sec:spinWaves}

Small-energy excitations of a spin system understood in terms of the
linearized dynamics around a fixed point are {\it spin waves.}  This
section is a study of the linearized dynamics around the ferromagnetic 
$(FM)$, antiferromagnetic $(AFM)$ and antiparallel
fixed points $(A).$  The term 'spin wave' is usually used to refer to
excitations of a ground state,  but the concept remains useful at points
such as the antiparallel fixed points which are saddles of the energy
function.

\par

The linearization of  \eqn{\ref{eq:newHamiltonian}} near the
the ferromagnetic (FM) fixed point is
\begin{equation}
H_{FM} \approx 3 - {3 \over 2} \left( \Phi_A^2 + \Phi_B^2 \right) 
- \left( {3-\sigma \over 2} \right) \left( Z_A^2+Z_B^2 \right).
\end{equation}
There are two degenerate spin waves with period
\begin{equation}
T_{FM}(\sigma) = { 2 \pi \over \sqrt{3(3-\sigma)} }.
\end{equation}
In the case of $\sigma=0.5,$ for which the quantum mechanics have been
extensively studied (see \cite{Houle98b}), the
analytic value of $T_{FM}=2.294$ agrees with the limit of the periods
of all fundamental orbits (computed by numerical integration,  see
\ref{sec:c3pos}) as
$E \rightarrow 3.$

\par

A similar expansion is possible around
either  AFM ground state ,  where $\Phi_A$ and 
$\delta_B=\Phi_B-{2 \sqrt{2} \pi \over 3}$  are small.
We obtain
\begin{equation}
H_{AFM}\approx  E_{AFM} - {3 \over 4} \left( \Phi_A^2 + \delta_B^2 \right) 
+ {1 \over 2} \left( Z_A^2+Z_B^2 \right).
\label{eq:afmExpansion}
\end{equation}
Here there are two degenerate spin waves with period
\begin{equation}
T_{AFM}(\sigma) = \sqrt{2 \over 3}{ 2\pi \over \sqrt{\sigma}}
\label{eq:afmsw}
\end{equation}
Approaching the isotropic case,  $\sigma \rightarrow 0$,  $T_{AFM}$ becomes
 infinite.  This coincides with the
exact solution for $\sigma=0$ in which all three spins precess around
the total spin vector \cite{Magyari87} at a rate proportional to the
length of the total spin vector: 
at $AFM$,  the total spin and the rate of spin precession are both zero.
When $\sigma=0.5$,  the periods of all fundamental
orbits converge to $T_{AFM}=7.255$ as $E \rightarrow - 1.5$ as
predicted by \eqn{\ref{eq:afmsw}}.  As the spin wave frequency drops to
zero,  the zero point energy of the quantum ground state also drops to
zero,  converging on the classical ground state energy as is observed in
\cite{Houle98b}.

\par

Although $\Mock_i$ is a saddle point rather than a ground state,  it is 
still possible to linearize the Hamiltonian in its vicinity.  Let
$\Phi_B$ and $\delta A$ be small,  where
$\delta A=\Phi_A-\bar{\Phi}_A,$ and $\bar{\Phi}_A=-\sqrt{2/3} \pi.$  Then,
\begin{eqnarray}
H_A &\approx& E_A
+ \left[ \left( { 1 \over 6 } + {\sigma \over 2} \right) Z_A^2 + {3
\over 2} \delta_A \right]\nonumber\\
&&\;\;\;\;\;\;\;\;\;\;\;\;\;\;\; -\left[ { (1-\sigma) \over 2 } Z_B^2 + { \Phi_B^2 \over 2 } \right].
\label{eq:mockEx}
\end{eqnarray}
The first set of terms in \eqn{\ref{eq:mockEx}} depends on $Z_A$ and $\Phi_A$ 
and the
second set depends on $Z_B$ and $\Phi_B.$
The first set
in \eqn{\ref{eq:mockEx}}
describes positive energy spin waves that live on the $Z_B=\Phi_B=0$
counterbalanced manifold with period
\begin{equation}
T_{A,c}(\sigma)={2 \pi \over \sqrt{1 + 3 \sigma}},
\end{equation}
which is $T_{A,c}=3.975$ when $\sigma=0.5.$  
The second set describes
negative energy spin waves that live on the $Z_A=\Phi_A=0$ stationary spin manifold
with period
\begin{equation} 
T_{A,ss}(\sigma)={2 \pi \over \sqrt{1-\sigma}},
\end{equation}
which is $T_{A,ss}=8.885$ when $\sigma=0.5.$  Corners of the
counterbalanced and stationary spin manifolds touch at right
angles at $\Mock.$  We will later take advantage of this
to compute the stability properties of the stationary spin and
counterbalanced orbits in Section \ref{sec:ccorbit}.


\Section{The topology of the energy surface}
\label{sec:c3topology}

Unlike the canonical cases of two-degree of freedom Hamiltonian
dynamics,  such as the Anisotropic Kepler Problem (AKP) and 
the Henon-Heiles Hamiltonian,  the three-spin system
exhibits nontrivial changes in the topology of the energy surface at
certain energies.  There are three transition energies: (i) the {\it
coalescence energy} $E_c(\sigma);$ (ii) the {\it antiparallel energy}
$E_A=-1;$ and (iii) the {\it polar energy} $E_p(\sigma)$.  Fig.
\ref{fig:topTransition} depicts the transition energies as a function
of $\sigma$ while Fig. \ref{fig:scenarioTwo} illustrates the
topology changes for $\sigma=0.5$.  As we increase
energy from the ground state $E=-1.5$, transition (i)
always occurs first.  If $0<\sigma<2/3$, transition (iii)
occurs before transition (ii),
otherwise when ${2/3}<\sigma<1$ (ii) occurs before (iii).

\subsection{The coalescence transition}
\label{sec:coalescence}

At low energies (near the AFM fixed points)
an energy barrier separates the two antiferromagnetic ground states.  Therefore, the energy surface is composed of two disconnected
parts.  Those parts become connected when $E=E_c(\sigma).$  
Using polar coordinates,
$\SPIN=(\sin \theta \cos \phi, \sin \theta \sin \phi, \cos \theta),$
the surfaces first touch at the saddle points on the $\phi=0$ line 
on Fig. \ref{fig:sphere} (one is hidden behind the sphere.) Thus, $E_c(\sigma)$
is found by considering \eqn{\ref{eq:ssSingle}},  the single-spin 
Hamiltonian for the stationary-spin manifold.  The saddle lies on the $\phi=0$ line, at the point where
${\partial H/ \partial \theta}=0$,  or
\begin{equation}
\sin \theta = { 1 \over 2-\sigma }.
\end{equation}
Substituting this back into \eqn{\ref{eq:ssSingle}},  the saddle-point
energy is
\begin{equation}
E_c(\sigma)={3 - 3 \sigma + \sigma^2 \over \sigma - 2}. \label{eq:afmcat}
\end{equation}

The coalescence occurs between (a) and (b) in
Fig. \ref{fig:scenarioTwo}.  An interesting quantum manfestation of
the coalescence transition was observed in \cite{Houle98b},  the tunnel
splitting of quantum levels as $E_c$ is approached from below.

\subsection{The antiparallel transition}

At $E_c$,  a set of necks come into existence that connect the
two lobes of the energy surface that were disconnected at energies below
$E_c$ (See
Fig. \ref{fig:scenarioTwo}b) when $E=-1$ these necks fuse,  changing
the connectedness of the energy surface again (See Fig.
\ref{fig:scenarioTwo}c.)  This is the {\it antiparallel
transition} --
the antiparallel fixed point $A_1$ (see Section \ref{sec:c3fpa}) 
lies in the center of the large hole in Fig. \ref{fig:scenarioTwo}b.
Two families of periodic orbits disappear at this transition,  including
one branch of stationary spin orbits approaching from $E<-1$ as well as
the counterbalanced orbits approaching from $E>-1.$ (One aspect of Fig.
\ref{fig:scenarioTwo}b is deceptive.  Being a three-dimensional cut out
of a four-dimensioanl space,  it fails to show two other pairs of
connecting necks that surround the $A_2$ and $A_3$ antiparallel fixed
points -- for a total of six necks.)

\subsection{The polar transition}

The system attains extreme energies (both minimum
and maximum) only when the spins lie in the equatorial plane.
As a result,  there is both a minimum and a maximum energy at which one spin 
can point at a pole,  which is a saddle point in the full phase space.
These are the upper and lower polar transition energies -- these
thresholds are found by pointing one spin,  say spin 1,
at the north pole and finding the maximum and minimum energy configurations.
Setting $Z_A=\sqrt{3/2},$  in
\eqn{\ref{eq:newHamiltonian}-\ref{eq:endNewH}} we get
\begin{eqnarray}
H&=&- {(1-\sigma) \over 2} \left( {3 \over 2}+Z_B^2 \right)
\nonumber\\
&&
\;\;\;\;\;+
{1 \over 4}\sqrt{9-20 Z_B^2+4 Z_B^4}
\cos \left(2 \Phi_B\right).
\end{eqnarray}
which has a maximum at $H_{max}={3 \over 4} \sigma$ and a minimum at
$H_{min}={3 \over 4} \sigma-{3 \over 2}.$ The polar transition 
occurs between panels (c) and (d) in Fig. \ref{fig:scenarioTwo}:  at
this point the energy surfaces touch the enclosing hexagonal prism,
forming a network of necks connecting the energy surface to itself.

\subsection{The classical density of states}
\label{sec:classicalDOS}

Changes in the topology of the surface section have 
an interesting effect on the classical and quantum densities of states.
The weighted area of the energy surface,
\begin{eqnarray}
\rho_c(E)=\int d\Phi_A d\Phi_B dZ_A dZ_B \delta(E-H(H,E)) 
\end{eqnarray}
is the {\it classical density of states},  since it is proportional to the
quantum density of states.  \cite{endnote1} Fig.
\ref{fig:classicalDOS} is a plot of the classical density of states for the
reduced system as a function in energy.  We observe two interesting
features:  first,  a discontinuity in the slope of $\sigma_c(E)$ at
the coalescence transition (This is (a) in Fig. \ref{fig:classicalDOS}.)
Second,  the density of states is apparently flat between the lower
polar transition $E_p(\sigma)$ and the antiparallel transition -- although
we don't have an analytic understanding of the flat spot,  numerical
evidence suggests that it is exactly flat.


\Section{Fundamental Periodic Orbits and Global dynamics}
\label{sec:c3pos}

This section presents the main results we've determined from 
numerical integration of the equations of motion: a map of the
fundamental periodic orbits of the three-spin system for and
\Poincare\ sections depicting the global dynamics of the system
for $\sigma=0.5$.
Like any chaotic system, the three spin system has an
infinite number of periodic orbits.  However, a few short period
orbits form the skeleton of the system's
dynamics.  Four of these are known; the {\it
stationary-spin} orbit, the {\it counterbalanced} orbit, the {\it
three-phase} and the {\it unbalanced} orbit.  Fig. \ref{fig:poet} plots
the energy-time curves of the four orbits for $\sigma=0.5.$
(Spin trajectories for the four orbit types
are visualized in Fig. 1 of \cite{Houle98b})

Sections \ref{sec:ssorbit} - \ref{sec:3porbit} discuss the
stationary-spin,  counterbalanced,  three-phase and unbalanced orbits
respectively.
Section \ref{sec:c3fmend} and \ref{sec:c3afmend} discuss global dynamics
near the ferromagnetic ($E \rightarrow 3$) and antiferromagnetic ($E
\rightarrow -1.5$) ends.  Section \ref{sec:symdyn} points out
how symmetries of the Hamiltonian manifest in its
classical dynamics.

\subsection{Stationary spin orbits}
\label{sec:ssorbit}

The stationary spin orbits are simple to study because any point on a 
stationary spin invariant manifold (see Section \ref{sec:manifolds}) lies 
on a stationary spin orbit.  As there are three stationary spin
manifolds,  there are three families of stationary spin orbits related by symmetry.

For a stationary spin orbit, one spin (say, spin 1) is stationary in the
equatorial plane, while the other two spins move in distorted circles, $180^\circ$
out of phase.  Fig. \ref{fig:sphere} is a plot of the trajectories of one
of the moving spins,  based on the single-spin Hamiltonian 
\eqn{\ref{eq:ssSingle}}.  In the range
$-1<E<3,$ each family of stationary spin orbits has a single branch;
trajectories on the single-spin sphere are concentric
distorted circles centered around the FM fixed point.  Between $E=-1$
and $E=E_c$, two branches of periodic orbits exist: the {\it outer branch}, still
centered around the $FM$ fixed point, and the {\it inner branch},  centered around
the $A$ fixed point.  Below the 
coalescence energy $E_c(\sigma)$, the orbits reorganize into a different
pair of branches (left and right), one centered around each antiferromagnetic ground state.

The stationary spin orbit runs
along the $Z_A=0$ seam on the slice of the energy surface visualized in
Fig. \ref{fig:scenarioTwo}.  Fig.
\ref{fig:scenarioTwo}d and c represent the case where $E>-1$ and
only one branch of the orbit exists.  In Fig. \ref{fig:scenarioTwo}b, 
the outer branch runs along the outside of the surface while the inner branch 
rungs along the inside of the hole in the
surface.  Finally in Fig. \ref{fig:scenarioTwo}a, left and right branches of the the
stationary spin orbit exist on two separate lobes of the energy surface.

The energy-time curve for the stationary spin orbits is seen in
Fig. \ref{fig:poet}.  Although one would expect that the periods of the left
and right
branches in the $E<E_c$ regime are the same (because they are related by reflection
symmetry,)  it's a bit surprising that the period $T(E)$ of
the inner and outer branches in the $E_c<E<-1$ is also the same.  Since the coalescence
separatrix intersects the stationary-spin manifold,  the period of the
stationary spin orbits goes to infinity as $E \rightarrow E_c$ from either 
side,  with an observable effect on the quantum mechanical orbit spectrum.\cite{Houle98b}

In the case of $\sigma=0.5,$  the stationary spin orbit is unstable for $E>-1.$
For $-1<E<E_c,$  the inner branch of the stationary
spin orbit is stable and the outer branch is unstable.  The left and right branches
are stable as $E \rightarrow -1.5$ but become unstable as the energy increases
and chaos becomes widespread (the orbit does momentarily regain its stability
near $E=-1.22.$)
Fig. \ref{fig:ssStab} is a plot of the stability parameter $\rho=\lambda+\lambda^{-1}$ versus
energy for $E<-1,$   where $\lambda$ and $\lambda^{-1}$ are
eigenvalues of the transverse stability matrix; $|\rho|<2$ for a stable orbit and
and $|\rho|>2$ an unstable orbit.  \cite{Lichtenberg92}

\subsection{Counterbalanced orbits}
\label{sec:ccorbit}

The counterbalanced orbit family is also easy to study because,  like the
stationary spin family,  it lives on an invariant manifold.
Counterbalanced orbits
exist only in the range $E>-1,$  and at least for $0<\sigma<2,$  the
counterbalanced orbit is always stable.  As is the case for counterbalanced
manifolds,  there are three counterbalanced orbits related by
the symmetry $\transop.$ 

The spin wave analysis of Section 
\ref{sec:spinWaves} can be applied to the stability properties of the
inner branch stationary spin and counterbalanced orbits in 
the limit $E \rightarrow 1$.
A counterbalanced orbit (approaching from $E>-1$) and
the inner branch of a stationary spin orbit touch at each
antiparallel fixed point.  Using the linearization around the
antiparallel fixed point \eqn{\ref{eq:mockEx}},  we can establish
that both orbits are stable,  and compute the limiting value of
the stability exponent $\rho$ at $E \rightarrow -1$ for both
orbits.

Because there are two distinct frequencies in the linearized
dynamics around the antiparallel fixed point, \eqn{\ref{eq:mockEx}},
dynamics in the vicinity of the antiparallel fixed point are
structurally stable and, for close enough energies, should be similar
to the linearized behavior.  Focusing attention on the
counterbalanced orbit, the degree of freedom orthogonal to the
counterbalanced orbit is the stationary spin orbit -- therefore the
counterbalanced orbit is stable as $E \rightarrow -1.$  In one circuit
of the counterbalanced spin orbit,  a slightly displaced trajectory winds
around the orbit at the frequency of the stationary spin spin wave.  The winding
number is the ratio of the periods of the two orbits,  or

\begin{equation}
\theta = 2 \pi {T_c \over T_{ss}} = \sqrt{{1 - \sigma} \over {1+3 \sigma}}.
\end{equation}
For $\sigma={1 \over 2},  \theta={2 \pi \over \sqrt{5}}.$  The stability
parameter $\rho=2 \cos \theta$ equals $-1.891,$  and agrees
with the result obtained by numerical integration
(see Fig. \ref{fig:cbStab}).  Repeating this analysis
for the stationary spin orbit,  we obtain $\rho \rightarrow 0.175$ in
agreement with Fig. \ref{fig:ssStab}.

\subsection{Three phase orbits}
\label{sec:3porbit}

The three phase orbits do not lie on an invariant manifold and thus
have richer behavior than the previous two families of orbits.
Unlike the stationary spin and counterbalanced orbits for which
$\avgdotphi=0,$  
three phase orbits can exhibit {\it precession},
a secular trend in $\Phi_0$ and
therefore can be periodic orbits of the reduced system but not the full system.
The three phase orbit undergoes a pitchfork birfucation at transition
energy $E_b$ ($E_b \approx -0.75$ for $\sigma=0.5$.)  Above $E_b,$  a
single branch of non-precessing orbits exists,  but below the bifurcation
three branches of three phase orbits exist:  a non-precessing unstable orbit
and two stable orbits for which
$\avgdotphi \neq 0$ with opposite signs.  This transition is visible 
on curve (c) of Fig. \ref{fig:poet}.  The unstable orbit exists for a 
small energy below $E_b$ but soon disappears when the spin trajectory
intercepts the poles of the spin spheres.

The three-phase orbits are so called because,  in three-phase orbits,  the 
spins each execute an identical circuit around a distorted
circle,  each $120^\circ$ out of phase -- much like the
currents used in three-phase AC power transmission.  The
multiplicity of the three-phase orbit is different from the previous two:
Above $E_b$ 
there are two three-phase orbits,  one in which the spins rotate clockwise
and another with counterclockwise rotation.  Below $E_b$ and the demise
of the nonprecessing orbit,  there are a total of four:  each AFM ground state
has it's own pair,  one member of which has $\avgdotphi$ positive and
the other $\avgdotphi$ negative.  Fig. \ref{fig:precSq} is a plot of
the per-orbit precession rate of the three-phase orbit below the pitchfork
bifurcation;  note that the precession rate converges on $2 \pi$ 
(effectively zero) as $E \rightarrow -1.5.$

The three phase family is more difficult to study than the previous two,
because we must search the energy surface for it.  It's still quite
straightforward,  for as seen in Sections
\ref{sec:c3fmend} and \ref{sec:c3afmend} the three spin
orbit lies on the $Z_B=0$ line of the $\Phi_A=0$ surface of section and
can be found by a one-dimensional search.  The stability exponent $\rho$ of
the three-spin orbit can be seen in Fig. \ref{fig:tpStab}.

\subsection{Unbalanced Orbits}

The {\it unbalanced} orbits,  a family of unstable periodic orbits, 
exist between $E=-1.5$ and $E=E_p.$   Unbalanced orbits do not lie on
a symmetric manifold and do not precess. The unbalanced family 
corresponds with the unstable fundamental periodic orbit of the \Henon-Heiles
problem near its ground state -- in the $(\Phi_A,\Phi_B)$ plane the projection
of an unbalanced orbits is roughly a parabola that does not pass through the
projection of the AFM fixed point.  Like the first two orbit
families,  the behavior of one spin in the unbalanced orbit is different
from the other two;  therefore there are three unbalanced orbits for each
of the two AFM fixed points,  for a total of six unbalanced orbits.
At low energies,  the odd spin moves along a closed curve in $(\phi,\theta)$ space while the
other two spins move along open curves which are dented on one corner and are
mirror images of one another.  One unbalanced orbit lies on the line $Z_B=0$ line of
the $Z_A=0$ surface of section.

The unbalanced orbit disappears at the polar transition, at which point
the trajectory of the odd spin grazes the poles, touching the holes
that appear between Fig. \ref{fig:scenarioTwo}c and 
\ref{fig:scenarioTwo}d.


\subsection{Dynamics near the ferromagnetic end}
\label{sec:c3fmend}

The periodic orbits are the ``skeleton'' of the dynamics of a system:
to understand the ``flesh'' requires the global view
obtained through \Poincare\ sections.  Choosing a good trigger plane
for our section was a matter of studying the 
projection of orbits in the $(\Phi_A,\Phi_B)$ plane:
to ensure that all
fundamental orbits appear in the \Poincare\ section,  our criteria
were that:  (i) all orbits crossed the trigger plane, and
(ii) no orbits were confined to the trigger plane.  $\Phi_A+\Phi_B=0$ satisfied both requirements.

($\Phi,Z$) coordinates improve the quality of our \Poincare\ sections 
compared to previous works on the three-spin system. \cite{Nakamura86a}
In previous works, \Poincare\ sections were taken with trigger
$d\spin_{1z}/ dt=0$ and projected on the $\spin_x$ and $\spin_y$
planes.  When this is done, the collective precession of the three
spins cannot be visually separated from more interesting degrees of
freedom.  Although the concentric loops of KAM tori can be seen in the
figures of \cite{Nakamura86a} when the trajectories on the tori are
non-precessing, they are superimposed by random dots from precessing
chaotic trajectories.  Worse, at energies close to the antiferromagnetic
ground state ($E=-1.5$), trajectories on KAM tori themselves precess,
destroying their image.  As a result, the sections of \cite{Nakamura86a} had
limited utility as a map of the dynamics of the three spin system and
had to be supplemented with power spectra of the classical
trajectories to determine if trajectories were regular or chaotic.

Fig. \ref{fig:c3fmp} is a surface of section using the $\Phi_A+\Phi_B=0$
trigger which we produced using a method of visualizing
\Poincare\ sections for two degree of freedom systems in three
dimensional space described in Appendix \ref{sec:c3killer}. 
Fig. \ref{fig:c3fmp} is a single
image of a simulated 3 dimensional object which can be interactively rotated 
and viewed from arbitrary positions.  The dark
opaque object is the $E=2.05$ energy surface,  which is approximately
an oblate spheroid with the $(Z_A,Z_B)$ plane passing through the equator.
Over that surface is plotted cloud of dots which
are the intersections of $E=2.0$ trajectories with the surface of
section. 

All of the fundamental periodic orbits intersect the
surface of section in two places, once passing through the surface of
section in the positive direction ($\dot{\Phi}_A>0$) and once in the
negative direction ($\dot{\Phi}_A>0$.)  Just on the lower visible edge of
the energy surface is a sort of terminator which divides
trajectories that cross the surface of section in the positive and negative
directions; this curve is not quite a geodesic but it does divide the energy surface into two approximate hemispheres.  The intersections
of the stationary-spin and counterbalanced orbits
with the surface of section form a ring of 12 fixed points lying in the $(Z_A,Z_B)$ plane with exact 12-fold 
symmetry while the two three-phase orbits cross the surface of section
away from the plane.

     The nature of KAM tori in the ferromagnetic limit $E \rightarrow 3$
is visible in Fig. \ref{fig:c3fmp}.  A concentric family of KAM tori exist
around each counterbalanced orbit,  and families of tori also exist
centered around the three-phase orbits.  Stationary spin orbits lie on the separatrix which divides counterbalanced
tori from three spin tori.  As energy is lowered,  this separatrix is
the first place where tori break and chaos is observed.

\subsection{Dynamics near the antiferromagnetic end}
\label{sec:c3afmend}

     To study the dynamics of the three-spin system near the
antiferromagnetic end,  $E\rightarrow -{3 \over 2}$  we chose
$\Phi_A=0$ as a trigger.  Although this violates criterion (ii) of
Section \ref{sec:c3fmend},  we gain the advantage that this trigger
plane extends from the lowest to the highest energies and crosses
both antiferromagnetic fixed points.  The practical disadvantage is that
a stationary spin orbit exists on the $\Phi_A=0$ line,  and appears as a
curve on the \Poincare\ section rather than a point,  but this does not
terribly complicate the interpretation of the section. 

Fig. \ref{fig:toChaos} illustrates the transition to chaos in the
antiferromagnetic regime with $\sigma=0.5$.  At $E=-1.39$ (see
Fig. \ref{fig:henonLike}) most tori are unbroken and
motion is primarily regular.  By $E=-1.35$ chaos is becoming noticeable in
separatrix regions,  and by $E=-1.3$ chaos is widespread.  At $E=-1.2$ no
islands of regular motion are obvious.  However, we know that regular
islands do exist because the inner branch of the stationary spin orbit is
stable at some energies in this regime (See section \ref{sec:c3pos}.)  The
transition to chaos on the antiferromagnetic side has been observed
previously \cite{Nakamura86a} in the same energy range.

\subsection{Symmetry and dynamics}
\label{sec:symdyn}

    Because the Hamiltonian (\ref{eq:newHamiltonian}) is threefold symmetric
around the antiferromagnetic fixed points,  (see Section \ref{sec:pzsym})
the low energy behavior of the
three spin system falls into the same universality class as the well-known
\Henon-Heiles system with Hamiltonian \cite{Gutzwiller90}
\begin{equation}
H={1 \over 2m} ( p_x^2+p_y^2 ) + { m \omega^2  \over 2} (x^2+y^2)
+ \lambda (x^2 y-y^3/3).
\end{equation}
This can be seen in Fig. \ref{fig:henonLike}, which looks remarkably like
\Poincare\  sections of the \Henon-Heiles system.  \cite{Gustavson66} 

Symmetries around a fixed point determine many properties of the dynamics
of a system in its vicinity including the nature of the fundamental orbits,
the global geometry of trajectories in phase space,  and degeneracies in
orbit frequencies.
Threefold rotation symmetry around the antiferromagnetic fixed
point ensures that the periodic orbits and KAM tori near one AFM fixed
point can be mapped 1-1 to those in  \Henon-Heiles,  but it also 
guarantees that the frequencies of all periodic orbits converge in the 
$E \rightarrow -1.5$ limit:  if we perform a Taylor series expansion
of the Hamiltonian at the fixed point (as in \eqn{\ref{eq:afmExpansion}}) 
the only quadratic term compatible with threefold rotation
symmetry is that with circular symmetry in the $(\Phi_A,\delta_B)$ and
$(Z_A,Z_B)$ planes.

The three-spin system exhibits a six-fold rotation symmetry near the
FM limit which is responsible for a different orbit and torus geometry in the
$E \rightarrow 3$ limit which is probably generic for Hamiltonian fixed
points with sixfold symmetry.  With six-fold symmetry,
the first and second derivatives of the time-energy 
curves are the same for all fundamental orbits at
$E=3;$ as a result,  the periods of orbits are
remarkably degenerate for a large range in energy (see Fig.
\ref{fig:poet}.)


\Section{Conclusion}
\label{sec:c3conclusion}

This paper is a detailed analysis of the classical dynamics of the
three spin cluster with Hamiltonian \eqn{\ref{eq:hamiltonian}}; many of the
features we find are connected with quantum phenomena in the accompanying
paper \cite{Houle98b}.
One class of phenomena are connected with changes in the
topology of the energy surface (see Section \ref{sec:c3topology}) which
occur as a function of energy: if we make a plot of quantum energy
levels as a function of $\sigma,$ shown in Fig. 5 of \cite{Houle98b}, we observe a tunnel splitting as pairs of
near-degenerate levels (at $E<E_c(\sigma)$) cross the $E_c(\sigma)$ curve (see
\eqn{\ref{eq:afmcat}}.)  This is caused by tunneling
between quantum states localized on the two disconnected parts of the 
energy surface.  A second
phenomenon related to the topology of the energy surface is that the
classical and quantum densities of states are apparently constant as
a function of energy for $E_c(\sigma)<E<-1$ (see Fig. 4 of \cite{Houle98b}
and Section \ref{sec:classicalDOS} of this paper.)

Our analysis of fundamental periodic orbits in Section
\ref{sec:c3topology} also has significance for the quantum problem.  The
Gutzwiller trace formula \cite{Gutzwiller90} predicts that classical
periodic orbits cause oscillations in the density of states.  In
\cite{Houle98b} we observed these oscillations by applying spectral analysis to the quantum density of states.
(see Fig. 3 of \cite{Houle98b}.)

Some technical aspects of the work presented in
this paper are interesting.  First, the change of coordinates that presented in
Section \ref{sec:change} enables us to understand the three-spin
cluster better than previous studies
\cite{Nakamura85} \cite{Nakamura93}
as we take advantage of the clusters conservation of total $\spin_z$
to reduce the dynamics to a tractable two-degree of freedom system.
Second, our use of three-dimensional visualization for
visualizing the energy surface and \Poincare\ sections 
clarifies
interpretation of \Poincare\ sections when the topology of the
energy surface is complicated.  Even in situations where the topology
is simple (such as is discussed in Section \ref{sec:c3fmend},)
three-dimensional visualization hides fewer symmetries of system than
the customary two-dimensional projection and eliminates the
confusion caused when two sheets of the energy surface are projected on top
of one another.  This method is described further in
Appendix \ref{sec:c3killer}

The three-spin cluster has similiaries to certain well-studied systems.
Our system has features in common with
the well-known Anisotropic Kepler Problem (AKP).
\cite{Gutzwiller90}.  Like the AKP,  total angular momentum around the
z-axis $\sum S_i^z$ is conserved, leaving two nontrivial degrees of
freedom.  Both the AKP and \eqn{\ref{eq:hamiltonian}} have a single
parameter ($\sigma$ in the case of our system) and are nonintegrable
for all values of the parameter save one ($\sigma=0$ in our case.)  In
both systems, all three classical frequencies are identical in the
integrable case.  For the AKP,  the integrable case is the ancient 
Kepler problem in which all trajectories are closed ellipses.  In our
problem,  in
the integrable case $(\sigma=0),$  all three spins precess around the total spin vector $\sum_i \SPIN_i$
at a rate proportional to the length of the total spin
vector. \cite{Magyari87} Our system is different from the AKP in a
number of ways.  First, the AKP is highly chaotic
throughout the parameter space in which is has been studied
\cite{Gutzwiller90} (The first stable periodic orbit was found after the
AKP had been studied for 14 years. \cite{Brouke85})  Our system, on
the other hand, shows highly regular behavior in much of the
parameter space (For instance, when $E>2$ in the $\sigma=0.5$
case) as well as irregular behavior in other areas (For instance,
$\sigma=0.5$ and $E \approx -1.2.$) Thus the elegant
application of symbolic dynamics to the AKP \cite{Gutzwiller90} is not
possible for our system.

Another connection between the three-spin cluster and a
well-studied system is the similarity between the dynamics of the
three-spin cluster in the antiferromagnetic limit and the \Henon-Heiles
problem.  \cite{Gustavson66,Henon64} The connection here is
most obvious in the surface of section shown in Fig. \ref{fig:henonLike} and occurs
because the three-spin cluster has a three-fold rotational symmetry around
the antiferromagnetic ground states similar to the symmetry of the
\Henon-Heiles problem.

In this work we have gotten a more intimate understanding of a
nonintegrable spin cluster than has been previously available supporting the
work described in  \cite{Houle98b},  which establishes
that periodic orbit theory can be applied to spin.  This work was funded 
by NSF Grant
DMR-9612304, using computer facilities of the Cornell Center
for Materials Research supported by NSF grant DMR-9632275.  We would
like to thank Masa Tsuchiya,  Jim Sethna,  and Greg Ezra for interesting
discussions.

\appendix

\Section{Numerical Integration in $\Phi,Z$ coordinates}
\label{sec:numint}

An important decision in the numerical study of the three spin problem
is the choice of variables to used to integrate the equation of motion.  This
choice affects the speed,  complexity and reliability of integration as
well as the range of \Poincare\ sections that can be easily taken. 

Our ODE integrator 
library was
written in Java and evolved from the software used for the
results published in \cite{Houle97}.  For both vector components and
$(\Phi,Z)$ coordinates we used adaptive fifth-order Runge-Kutta
integration based on the code from \cite{Press92} although our system
allows the use of different integrators such as fourth-order fixed
Runge-Kutta for testing.  \cite{endnote2}

In the early phase of this work we integrated the spin vector components
$(\spin_x,\spin_y,\spin_z)$ of the individual spins.
The vector component representation has several advantages:  
the software requirements are simple and
it's straightforward to write a general routine for 
evaluating the equations of motion for any spin Hamiltonian which is
polynomial in $\spin_x$, $\spin_y$ and $\spin_z.$  Spin vector coordinates
are also free of obnoxious singularities.  However,  the need to
isolate the overall precession of the spins from more interesting motions
led us to integrate the system in $(\Phi, Z)$ coordinates so we could
easily set \Poincare\ sections in the $\Phi$-space. (the
rationale for setting triggers in $\Phi$ space is discussed in Section
\ref{sec:c3fmend}.)  

Although the transformation $(\phi,z)
\rightarrow (\Phi,Z)$ is a straightforward linear transformation, the
need to use inverse trigonometric functions to convert $\SPIN$ into
$(\phi,z)$ adds overhead and, more seriously, additional complexity
to deal with branch cuts.  (Numerical algorithms that work
with branch cuts, particularly involving square roots, are
difficult to design.  Failure modes caused by roundoff error 
with a probability of $10^{-6}$ per dynamical timescale are a major
complication for a program that calculates thousands of trajectories.) 
Simple strategies for disambiguating branch cuts that do
not introduce an error-prone memory between steps lose
valuable topological information.  If, for instance, $\Phi_0$ is
computed from some manipulation of the $\SPIN_i$ components, and, say,
is always in the range $0<\Phi_0<{4 \pi \sqrt{3}}$ it isn't as easy
to determine the net precession (secular trend of $\Phi_0$) of a periodic orbit
as it would be if $\Phi_0$ were integrated directly.

We performed all of the integrations in this work in $(\Phi,Z)$ coordinates,  
with equations of motion derived
from \eqn{\ref{eq:newHamiltonian}}.  The main difficulty we had is that
the integration can fail on a trajectory on which a spin passes through a
pole;  this is not a failing of the $(\Phi,Z)$ coordinates as much as 
of the $(\phi,z)$ coordinates.  As spin $i$ passes close to the pole,
the singularity in the mapping from $(\phi_i,z_i) \rightarrow \SPIN_i$ forces
$\dot{\phi}_i \rightarrow \infty.$  If the trajectory misses the pole by
more than $1e-6$ in the $z$ axis when using double precision math,  the
primary consequence is that the adaptive step size integrator reduces the time
step and  integration is slowed.  If the trajectory passes
much closer to the pole,  however,  no step size may have a sufficiently
small error estimate and the adaptive step size algorithm will fail.
Although it would be possible
to avoid this problem by either switching to spin vector coordinates when
the trajectory passes close to the pole or by adding more intelligence (and
possibly bugs) to the adaptive step size algorithm,  in practice it affects a
small enough volume of phase space that it only manifests
when investigating trajectories specifically chosen to 
pass near a pole.

Sometimes it is necessary to work with ($\Phi_A$,$\Phi_B $) normalized
to the unit hexagon, for instance, to set a \Poincare\ section trigger
on $\Phi_A=0.$ At energies above the upper polar transition and below
the lower polar transition (see Section \ref{sec:c3topology}) this is
not necessary because the trajectory does not wander long distances in
the ($\Phi_A$,$\Phi_B$) plane.  For some time this prevented us from
taking \Poincare\ sections in the region between the two transitions,
since the trajectory would eventually wander far from the trigger
plane.  To solve this,  we found an algorithm for mapping
$(\Phi_A,\Phi_B)$ back to the unit hexagon: first, (i) use the
modulus function to map points into the primitive
cell of the hexagonal lattice,  a rhombus.  Then, (ii) apply a unit vector
translation to those points that fall on corners of the rhombus
outside the unit hex to bring them into the unit hex.  


\Section{Solid \Poincare\ Sections}
\label{sec:c3killer}

\par

In the process of studying the three spin problem,  we found conventional methods of
drawing \Poincare\  sections inadequate and improved upon them by developing
a method for rendering \Poincare\  sections in three-dimensional
space.    This greatly simplified the interpretation of
\Poincare\  sections for our system.  Although not all systems pose as 
serious technical problems as ours,  we believe that this method clarifies
the geometry of \Poincare\ sections and can simplify
the presentation of \Poincare\ sections to audiences which are not
specialized in dynamics.  The techniques described in this appendix were used
to generate Fig. \ref{fig:scenarioTwo}, Fig. \ref{fig:c3fmp} and Fig. \ref{fig:toChaos}.

For Hamiltonian systems with two degrees of freedom,  the intersection of
the energy surface with the surface of section is a two-dimensional surface
embedded in a three-dimensional space (the surface of section).  Often,  this
intersection has the topology of a sphere -- this is true of
the \Henon-Heiles system as well as for the three spin systems above the
upper polar threshold and below the coalescence energy $E_c(\sigma).$  The
trajectory crosses the surface of section in two directions,  which we will
call the positive and negative directions.  Over part of the sphere,  the
trajectory crosses in the positive direction and over the rest of the sphere,
the trajectory crosses in the negative direction.  In between there is a 
seam over which the trajectory is tangent to the surface of section.

Difficulties arise when plotting the intersection of trajectories with the
surface of section even when the topology is simple.  To take a specific example,
consider the case of the three spin problem with $\sigma=0.5$ and $E=-2$ with
the trigger on $\Phi_A+\Phi_B=0.$  A reasonable set
of $(x,y,z)$ variables for this problem is $(x,y,z)=(Z_A,Z_B,\Phi_A).$
Conventional choices of projection are $(x,y)$ which would overlap the two
distinct three phase orbits and place the stationary spin orbits on the
edge of the plot,  or $(x,z)$ which obscures the symmetry of the stationary
spin and counterbalanced orbits seen in the $(x,y)$ plane.

\par

While writing the paper \cite{Houle97} we began development of a set
of Java class libraries for integrating differential equations and taking
\Poincare\  sections.  Interchangeable trigger modules are connected to
the differential equation integrator,  allowing the user to
set a triggering criteria for surface of section.  When the surface is
crossed,  we solve for the crossing time by varying the integration step
$\delta t$ and solving for the value of $\delta t$ that intersects
the section by the Newton-Raphson method.  (We also tried the method of
method of \Henon\ \cite{Henon82} in which a change of variable is made to
make the trigger coordinate become a time coordinate -- we found
that performance of our method and \Henon's method is similar,  but ours 
was more robust)
We primarily integrate with an adaptive fifth-order Runge-Kutta implementation,
but the design of
our program allows us to replace our integrator with another
fixed or adaptive step algorithm.

\par

One component of a 3-d \Poincare\  section is a rendering of the energy
surface.  For an arbitrary system,  the energy surface can be rendered by
treating it as an implicit function;

\begin{equation}
E(x,y,z)=E_0
\end{equation}

this sort of surface can be rendered using ray-tracing techniques,  or can
be rendered in a primitive manner by dividing the 3-space into voxels 
and coloring in cubes which are above or below a threshold value. 
A better method is to generate a polygonal mesh
with Bloomenthal's algorithm \cite{Bloomenthal88}.  The relative merits of various methods for
polygonizing a mesh are discussed in \cite{Glassner90}.  For a uniform mesh (not adaptively sampled),
Bloomenthal's algorithm is equivalent to a table-driven method known as {\it Marching Cubes}
\cite{Lorensen87}.  We can compute a mesh for our system in under 20 seconds and 2 megabytes of storage,  and this time could be reduced by the use of
Marching Cubes.  The resulting mesh can be rendered with a standard 3-d system,
either
an interactive system such as
as OpenGL or VRML or an off-line system such as POVRay or Renderman.

The generation of the actual \Poincare\ sections is quite straightforward.
We merely plot a set of (x,y,z) points when the trajectory intersects the
surface of section.  The last problem is determining initial conditions for injection.  In some cases we wish to
interactively choose injection points, in which case it's necessary to
translate a hand-chosen injection point in screen cordinates into
into a point one on the energy surface;  this way, in an interactive 
3-d environment,  a user can click on the  energy surface to set an injection point.  For most cases,  we prefer to have the computer automatically 
generate \Poincare\ sections,  choosing injection points that reveal all major
phase space structures.  To accomplish this,  our program divides the surface
of section into voxels (tiny cubes) and inspects the value of the energy
at the corners of the cubes to find which cubes intersect the energy surface.
Next,  the program scans the surface,  ensuring at least one trajectory
either originates in or passes through each voxel that contains part of the
energy surface.  To inject in a voxel,  our program computes
the gradient of the Hamiltonian at the center of the voxel and precisely
locates the energy surface by a Newton search along the line of fastest
change in energy.  It then marks that voxel as filled.  The trajectory is
then evolved forward in time for a fixed number of intersections:  at each
intersection,  the containing voxel is set as filled.  The program continues
to find voxels which are not filled and injects into them until all voxels
are filled.

Although the energy surfaces are interesting in themselves for systems
with topology changes (such as the three spin system),  they are also
essential for interpreting solid \Poincare\ sections.  To make a solid
\Poincare\ section we render a cloud of dots in three-dimensional space
at the points where trajectories intersect the surface of section.  The
resulting dot cloud is transparent and is difficult to make sense
of without an opaque object underneath it.  Several factors conspire
to confuse the eye.  For instance,  a part of a dot cloud that is
further from the viewer is made visually denser by perspective which makes
it appear heavier and more solid to the viewer -- causing the viewer to
conclude that the distant part of the dot cloud is closer.  An opaque
rendering of the energy surface eliminates the confusion caused when
images of the energy surface overlap.

\par

Because integrating differential equations is slow,  it takes about two hours
to compute our \Poincare\ sections.   Because the calculation can be broken
into a a number of mainly unrelated calculations,  parallelizing the
calculation across multiple processors on an SMP machine or a cluster of
computers would be straightforward and worthwhile.  

\par

A complicated aspect of the three-spin problem is the hexagonal tiling
of the phase space.  This introduced two
difficulties which required special solutions.  We implemented these
solutions using object-oriented techniques to replace general-purpose
classes with specialized subclasses.
First,  the 3-space $(Z_A,Z_B,\Phi_A)$ is bounded by the points
at which the spins touch the north pole.  The volume inside this boundary is
a hexagonal prism (See Section \ref{sec:howtile} and Fig. 
\ref{fig:scenarioTwo})
Using a cubical
grid to represent a non-cubical domain forced us to keep track of which
points were inside the hexagonal prism and which were outside.  Not only
was this difficult and slow,  but the irregular pixilation of the boundary
confused both the surface-constructing algorithms and the eye of the viewer,
resulting in surfaces with a jagged,  serrated appearance.  We solved the
problem by replacing the Java class responsible for mapping integer voxel numbers
to ($x,y,z$) coordinates with one that smoothly maps a 
cubic grid on an integer lattice to a hexagonal prism.  All sample
points fell within the allowed range.  We headed off numerical problems
caused by square roots in the energy function by multiplying
x and y coordinates by $1-10^{-13}$ before computing the energy.

\par

At energies between the upper and lower polar thresholds,  another 
set of problems arises.  Here,  the projection of a trajectory in ($\Phi_A,\Phi_B$)
can wander outside the unit hexagon.  If the trajectory wanders
in the ($\Phi_A,\Phi_B$) plane it may fail to intersect the
surface of section.  This problem
can be solved by wrapping the Java class which
provides for inspection and manipulation of the
system state in $(\Phi,Z)$ coordinates (the actual integrator was written
before these coordinates were settled) with a Proxy \cite{Gamma95} class which
maps $(\Phi,Z)$ coordinates back to the unit hexagon -- an algorithm for
performing this mapping is described in Appendix \ref{sec:numint}.

\bibliographystyle{h-physrev}
\bibliography{journals,misc,classical,semiclassics,magnetism,math,anons,informatics}

\begin{thebibliography}{27}
\bibitem{endnote2} A category of integrator
known as symplectic integrators preserve the symplectic property
of classical mechanics \cite{McLachlan92} and are claimed to be more
qualitatively accurate for very long integration times than
conventional integrators.  Because we were only interested in
short-term time evolution and because there is no systematic
procedure for deriving a symplectic integrator for an arbitrary
problem, we did not make the investment of developing a symplectic
integrator for our system.
\end{thebibliography}

\eject

\begin{figure}
\centerline{\epsfxsize=3.2in\epsfbox{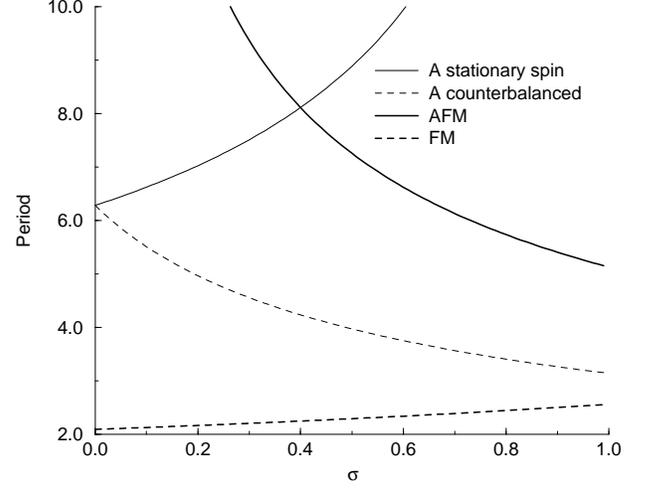}}
\caption{Spin wave frequencies as a function of $\sigma.$}
\label{fig:spinWaves}
\end{figure}
\eject

\begin{figure}
\centerline{\epsfxsize=3.2in\epsfbox{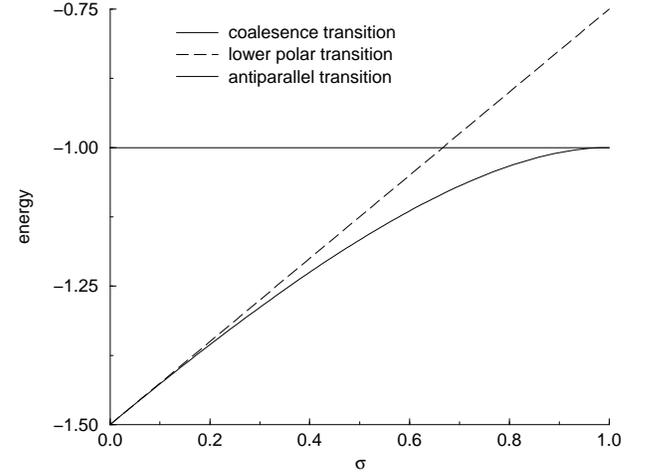}}
\caption{Energies at which transitions in the energy surface topology occur.}
\label{fig:topTransition}
\end{figure}
\eject

\begin{figure}
\centerline{\epsfxsize=3.2in\epsfbox{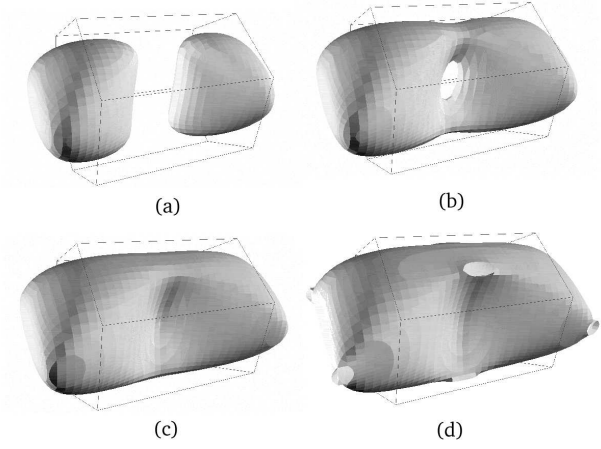}}
\caption{Changes in the topology of the energy surface
($H(Z_A,Z_B,\Phi_B)=E$ with $\Phi_A=0$),
for $\sigma> 2/3,$ 
here $\sigma=0.75$.  (a)$E=-0.92,$  (b) $E=-=0.98$, (c) $E=-1.02$, and (d) $E=-1.1.$ The
hexagonal prism represents the hexagonal boundaries of the $(Z_A,Z_B)$
plane: the horizontal direction is $Z_A$, the vertical direction is
$Z_B$ and the long axis of the prism is $\Phi_B.$ The endpoints
of the prism are at
$\Phi_B={2\sqrt{2}\over 3} \pi$ and $\Phi_B={4\sqrt{2} \over 3}\pi.$}
\label{fig:scenarioTwo}
\end{figure}

\eject

\begin{figure}
\centerline{\epsfxsize=3.2in\epsfbox{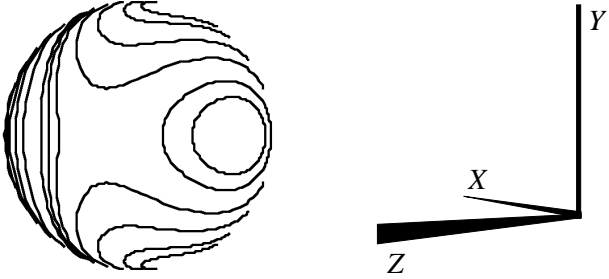}}
\caption{The trajectory of a moving spin on the stationary spin manifold, for
$\sigma=0.5.$
The well around the negative z-axis is the antiparallel fixed point,
while the wells above and below it are the two antiferromagnetic
fixed points.  (The FM fixed point is on the positive x-axis and is on the hidden
side of the sphere.)  The separatrix between orbits centered on the AFM fixed
points those centered around the antiparallel fixed points occurs at 
energy  $E_c$ (see \eqn{\ref{eq:afmcat}}.)} 
\label{fig:sphere}
\end{figure}
\eject

\begin{figure}
\centerline{\epsfxsize=3.2in\epsfbox{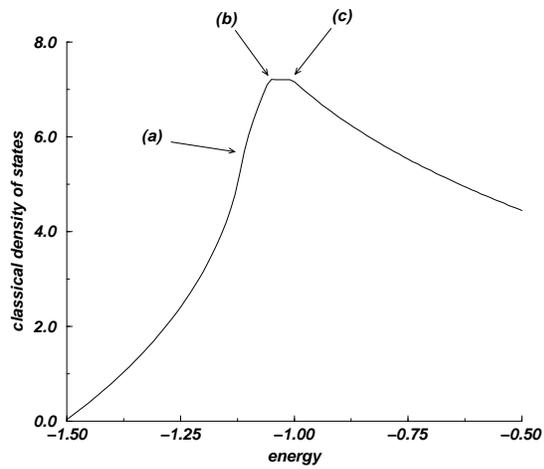}}
\caption{
The classical density of states as a function of energy for $\sigma=0.6$
determined by numerical integration.
(a) the coalescence transition,  (b) the lower polar transition,  (c) the
antiparallel transition.  The total area under the curve is $3 \pi,$  the volume
of the reduced phase space.}
\label{fig:classicalDOS}
\end{figure}
\eject

\begin{figure}
  \centerline{\epsfxsize=5in\epsfbox{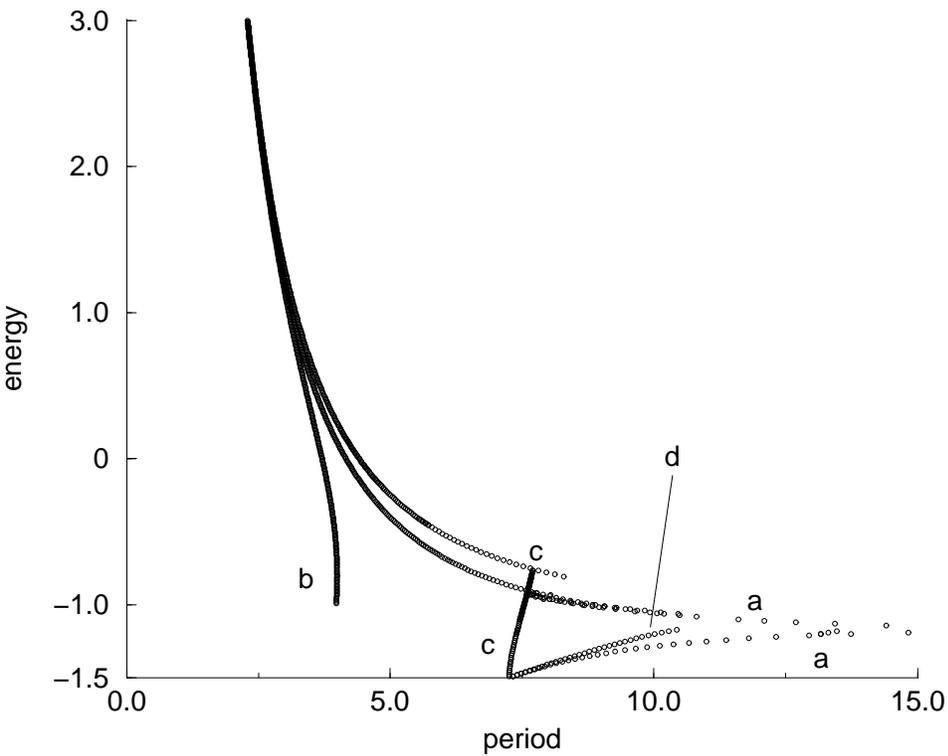}}
  \caption{
    Energy-period curve of the three spin system with $\sigma=0.5.$
    (a) Stationary Spin,  (b) Counterbalanced,  (c) Three-phase orbit,  and (d)
Unbalanced orbits.}
  \label{fig:poet}
  \label{fig:classicalEtau}
\end{figure}
\eject

\begin{figure}
\centerline{\epsfxsize=3.2in\epsfbox{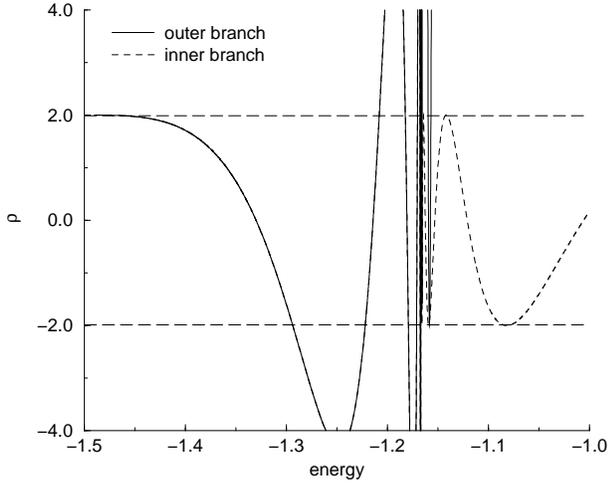}}
\caption{
The stability exponent $\rho$  ($\rho=\trace \monod$ where $\monod$ is the
stability matrix) of the stationary spin orbit as a function
of energy for $\sigma=0.5.$  An orbit is stable when $|\rho|<2$ and
unstable when $|\rho|>2.$  The stationary spin orbit is stable
near the AFM ground state and alternates between being stable
and unstable until $E_c.$  The outer branch of the orbit is
unstable for all $E>E_c$ while the inner branch appears to always be
stable until its disappearance at $E=1.$
\label{fig:ssStab}
}
\end{figure}
\eject

\begin{figure}
\centerline{\epsfxsize=3.2in\epsfbox{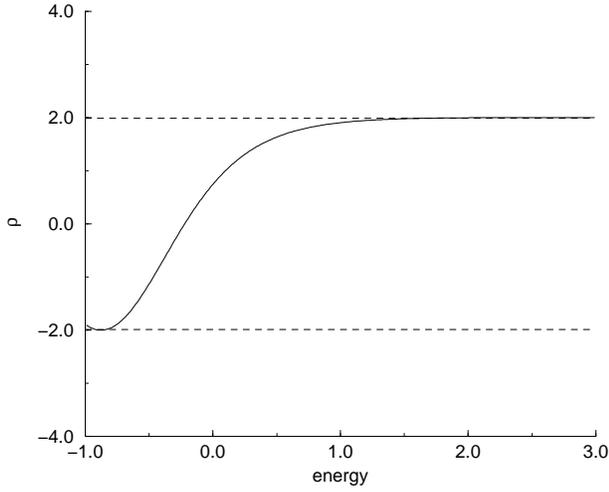}}
\caption{
The stability exponent $\rho=\trace \monod$ where $\monod$ is the
stability matrix for the counterbalanced orbit in the case
$\sigma=0.5$ where the orbit is always stable.
}
\label{fig:cbStab}
\end{figure}
\eject

\begin{figure}
\centerline{\epsfxsize=3.2in\epsfbox{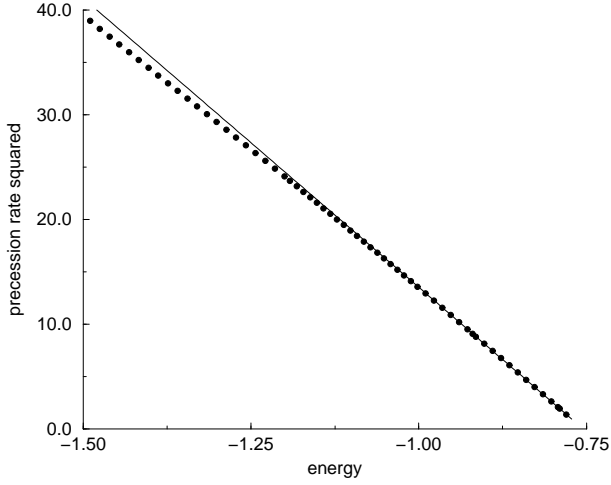}}
\caption{
The precession rate squared as a function of energy for the three phase
orbit with $\sigma={1 \over 2}.$
The precession rate is the
amount that $\Phi_0$ increases after one period of the orbit in the
reduced space.  In the limit of $E \rightarrow - {3\over2}$ the precession
rate becomes $2 \pi.$  The straight line fit (fit to
all of the points with $E > 0.90$) indicates $(E-E_b)^{1 \over 2}$ scaling
with $E_b=-0.7547.$  Note that $(2 \pi)^2\approx 39.48.$
}
\label{fig:precSq}
\end{figure}
\eject

\begin{figure}
\centerline{\epsfxsize=3.2in\epsfbox{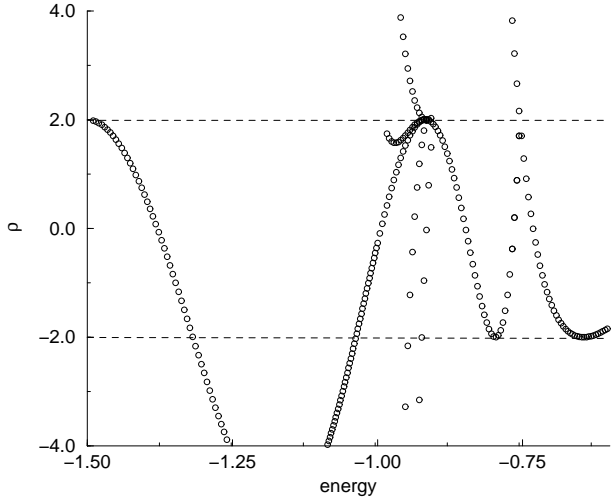}}
\caption{
Stability exponent $\rho$ for the three phase orbits for $\sigma=0.5$.
The three phase orbit is stable for $E>E_b,\  E_b\approx 0.75.$  Starting
from high energies, $\rho$ reaches $2$ at $E_b$ where a bifurcation gives birth to two stable
precessing orbits and an unstable nonprecessing orbit.  Another
(unstudied) bifurcation happens around $E=0.91.$
Starting at the low-energy AFM end,  the orbit starts out stable
and becomes unstable at $E=-1.32$ in a period doubling bifurcation.}
\end{figure}
\label{fig:tpStab}
\eject

\begin{figure}
\centerline{\epsfxsize=3.2in\epsfbox{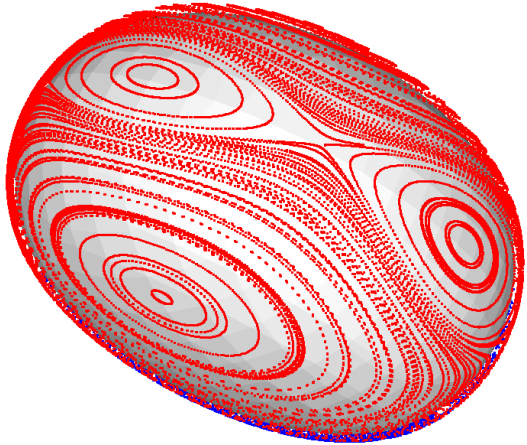}}
\caption{\Poincare\ section at $\sigma=0.5, E=2.0$ with a
$\Phi_A+\Phi_B=0$ trigger:  point (a) is a
stationary spin orbit,  point (b) is a counterbalanced orbit,
and point (c) is a three-phase orbit.  The energy surface is approximately
an oblate spheroid with stationary spin and counterbalanced
orbits on the equator on which $\Phi_B=0;$  the back side of the
surface is hidden.  Just behind the equator is the other
three-spin orbit for which the spins rotate opposite the
direction of the visible three-spin orbit.  Throughout most of the visible
hemisphere,  orbits cross the section traveling in the positive
direction ($\dot{\Phi}_A>0$);  on the invisible hemisphere,  the
picture is mirror-reflected across the ``terminator''
for orbits crossing in the negative direction.}
\label{fig:c3fmp}
\end{figure}
\eject

\begin{figure}
\centerline{\epsfxsize=3.2in\epsfbox{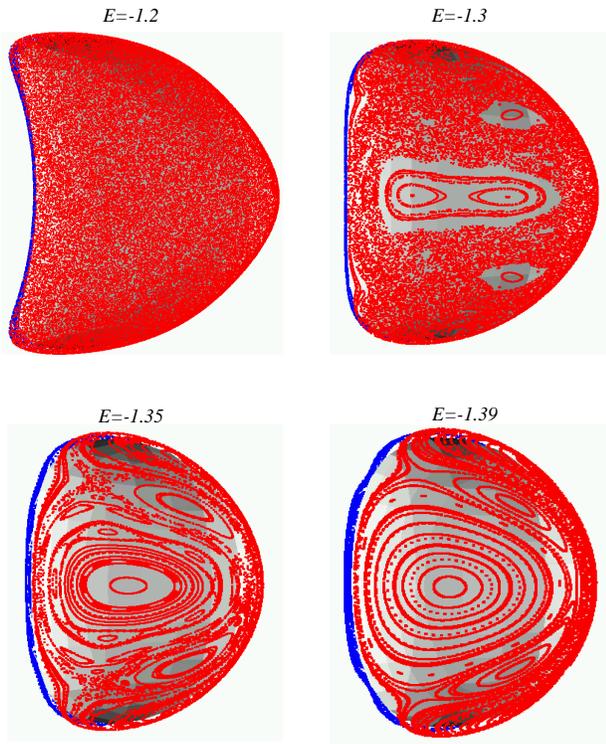}}
\caption{The transition to chaos at the antiferromagnetic end
($E \to -1.5$)
with $\sigma=0.5.$  
Shown is one of the two lobes from
Fig. \ref{fig:scenarioTwo}.  See Fig. \ref{fig:henonLike}
to determine which orbits are which.
}
\label{fig:toChaos}
\end{figure}
\eject

\begin{figure}

\centerline{\epsfxsize=3.2in\epsfbox{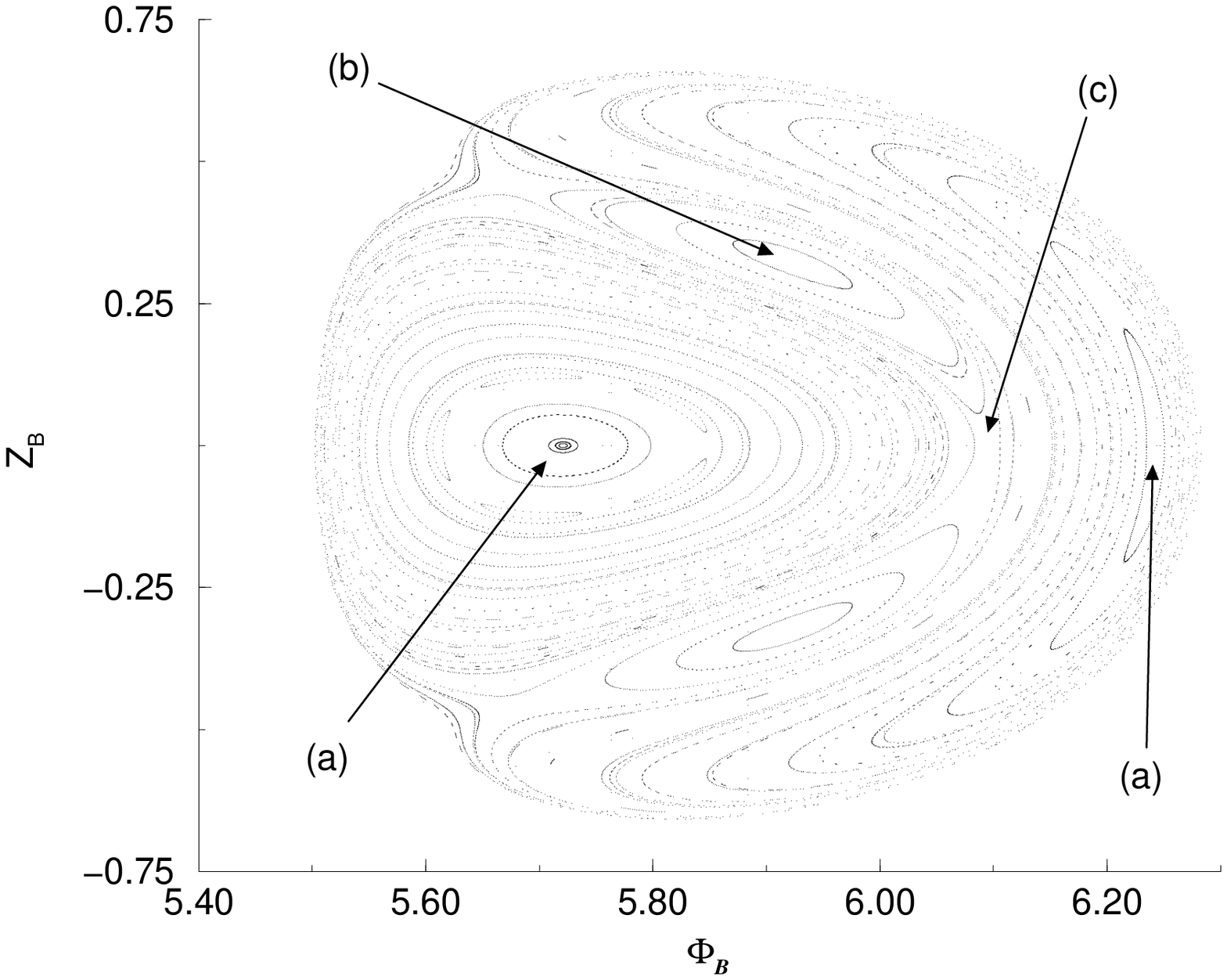}}
\caption{
A projection of the $E=-1.39$ \Poincare\ section
into the $(\Phi_B,Z_B)$ plane.  
Orbits marked
(a) are three phase orbits,  (b) is a stationary spin orbit,
and (c) is an unbalanced orbit.}
\label{fig:henonLike}
\end{figure}
\eject


\end{document}